\documentclass[aps,twocolumn]{revtex4}
\usepackage{amsfonts}
\usepackage{amsmath}
\usepackage{amssymb}
\usepackage{graphicx}

\begin{document}
\preprint{ }
\title{Paramagnetic anisotropic magnetoresistance in thin films of SrRuO$_{3}$}
\author{Isaschar Genish}
\author{Yevgeny Kats}
\author{Lior Klein}
\affiliation{Department of Physics, Bar-Ilan University, Ramat-Gan 52900, Israel}
\author{James W. Reiner}
\altaffiliation{Present address: Department of Applied Physics,
Yale University, New Haven, Connecticut 06520-8284}
\author{M. R. Beasley}
\affiliation{T. H. Geballe Laboratory for Advanced Materials, Stanford University,
Stanford, California 94305}

\begin{abstract}
SrRuO$_{3}$ is an itinerant ferromagnet and in its thin film form
when grown on miscut SrTiO$_{3}$ it has $T_{c}$ of $\sim150$ K and
strong uniaxial anisotropy. We measured both the Hall effect and
the magnetoresistance (MR) of the films as a function of the angle
between the applied field and the normal to the films at
temperatures above $T_{c}$. We extracted the extraordinary Hall
effect that is proportional to the perpendicular component of the
magnetization and thus the MR for each angle of the applied field
could be correlated with the magnitude and orientation of the
induced magnetization. We successfully fit the MR data with a
second order magnetization expansion, which indicates large
anisotropic MR in the paramagnetic state. The extremum values of
resistivity are not obtained for currents parallel or
perpendicular to the magnetization, probably due to the crystal
symmetry.

\end{abstract}




\maketitle

\section{Introduction}

The phenomenon of anisotropic magnetoresistance (AMR) in magnetic
conductors expresses the dependence of the resistivity $\rho$ on
the angle $\delta$ between the current $\mathbf{J}$ and the
magnetization $\mathbf{M}$. In polycrystals the AMR effect is
commonly found to follow: $\rho=\rho_{\perp
}+(\rho_{\parallel}-\rho_{\perp})\cos^{2}\delta$ where
$\rho_{\perp}$ is the resistivity when $\mathbf{J}\bot\mathbf{M}$,
and $\rho_{\parallel}$ is the resistivity when $\mathbf{J\Vert M}$
\cite{AMR}. This simple relation is not expected to hold in
crystalline samples where both the current orientation relative to
the lattice as well as the magnetization orientation relative to
the lattice play an important role.

Here we present AMR measurements of thin films of the $4$\emph{d}
itinerant ferromagnet SrRuO$_{3}$ above $T_{c}$ ($T_{c}\sim150$
K). Those films are epitaxial and characterized by large uniaxial
magnetocrystalline anisotropy (MCA) \cite{easy axis}. In our
measurements, we study the AMR in uncommon conditions: (a) while
in most AMR measurements the orientation of $\mathbf{M}$ is
changed without changing its magnitude, here, because of the large
MCA both orientation and magnitude of $\mathbf{M}$ are changing;
and (b) while in most AMR measurement the applied field
$\mathbf{H}$ is parallel to $\mathbf{M}$, here, because of the
large MCA $\mathbf{M}\nparallel\mathbf{H}$ except for the cases
where $\mathbf{H}$ is along the easy or hard axes. Therefore, to
explore the AMR in SrRuO$_{3}$ it is not sufficient to measure
magnetoresistance (MR) as a function of angle, but we need to
independently determine both the magnitude and orientation of
$\mathbf{M}$.

Our films are grown on miscut ($2^{\circ}$) SrTiO$_{3}$ substrates
using reactive electron beam epitaxy. The films have orthorhombic
structure ($a=5.53$ \AA , $b=5.57$ \AA , $c=7.85$ \AA )
 and they grow uniformly (without twinning) with
the $c$ axis in the film plane and the $a$ and $b$ axes at
$45^{\circ}$ out of the plane (see Fig. \ref{sam})
\cite{Marshall}. The MCA is uniaxial with the easy axis along the
$b$ axis and thus the uniform growth of the measured films can be
confirmed by MR measurements in the ferromagnetic phase
\cite{angle}. The films were patterned by photolithography to
allow Hall effect (HE) and MR measurements. The sample whose
results are presented here is $300$ \AA \ thick with $T_{c}$
$\sim147$ K and resistivity ratio of $\sim13$.

\section{Measurements and discussion}

Our measurements have two parts: (a) extraordinary Hall effect
(EHE)\ \cite{EHE} measurements from which we extract both the
magnitude and the orientation of $\mathbf{M}$, and (b) MR
measurements for currents in the $[001]$ and $[1\overline{1}0]$
directions. Combining the two measurements we show that the MR
can\ be fit very well with a second order magnetization expansion.

\begin{figure}[ptb]
\includegraphics[scale=0.35, trim=150 350 200 -180]{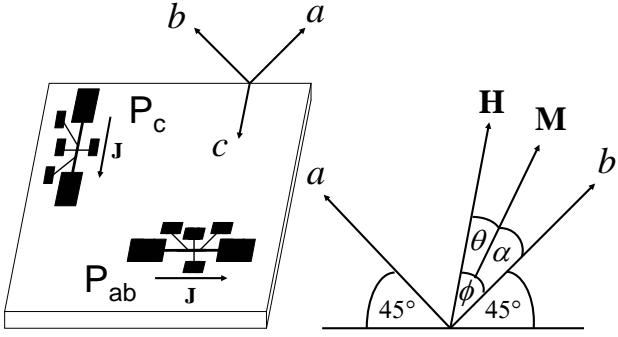}
\caption{A sketch of the patterned film. In pattern P$_{c}$ the
current $\mathbf{J}$ is in the [001] direction. In pattern
P$_{ab}$ the current $\mathbf{J}$ is in the $[1\overline{1}0]$
direction. The crystallographic directions $b$ (easy axis), and
$a$ (hard axis) are at 45$^{\circ}$ out of the plane of the film.
In our measurements the field $\mathbf{H}$ is rotating in the
(001) plane, $\phi$ is the angle between $\mathbf{H}$ and $b$,
$\alpha$ is the angle between the induced $\mathbf{M}$ and $b$,
and $\theta$ is the angle between $\mathbf{M}$ and $\mathbf{H}$.}
\label{sam}
\end{figure}

Both HE and MR measurements were performed as a function of the
angle $\phi$ between the applied field $\mathbf{H}$ and the easy
axis, where $\mathbf{H}$ is rotating in the $(001)$ plane. Each
film has two kinds of patterns: a pattern with current along the
$[1\overline{1}0]$ direction (denoted\ P$_{ab}$) and a pattern
with current along the $[001]$ direction (denoted\ P$_{c}$). While
in P$_{ab}$ the angle between $\mathbf{J}$ and $\mathbf{H}$ varies
with $\phi$, in P$_{c}$ the field $\mathbf{H}$ is always
perpendicular to $\mathbf{J}$. The measurement configuration is
illustrated in Fig. \ref{sam}.

\begin{figure}[ptb]
\includegraphics[scale=0.5, trim=150 0 200 0]{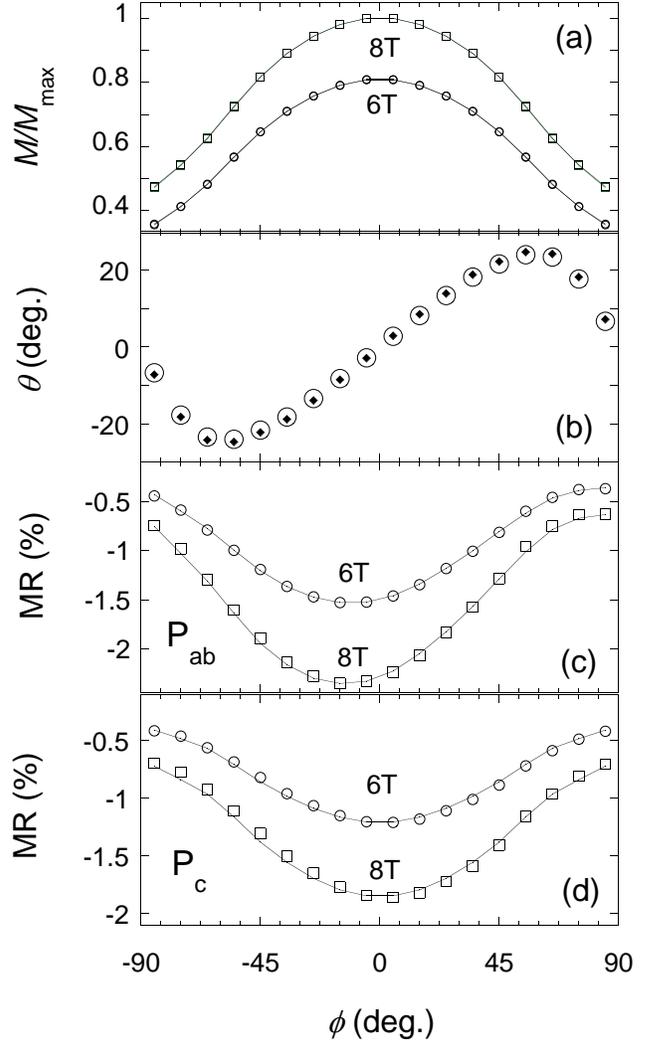}
\caption{Magnetization and MR data at $170$ K as a function of the
angle $\phi$ between $\mathbf{H}$ and the easy axis $b$, and fits
of the MR data based on the measured $\mathbf{M}$ and Eq. \ref{MR
expansion}. (a) The magnitude of $\mathbf{M}$ relative to
$M_{\max}$, which is the magnitude of $\mathbf{M}$ obtained with
$H= 8$ T applied along the easy axis. (b) The angle between
$\mathbf{M}$ and $\mathbf{H}$ for $H=8$ T (empty circles) and
$H=6$ T (full diamonds). (c) MR in P$_{ab}$. (d) MR in P$_{c}$.
For graphs (c) and (d) the points represent the experimental
results and the lines represent the fit.} \label{mr}
\end{figure}

The Hall field in magnetic conductors has two contributions:
\[
\mathbf{E}_{H}=-R_{0}\mathbf{J}\times\mathbf{B}-R_{s}\mathbf{J}\times\mu
_{0}\mathbf{M}
\]
where $\mathbf{B}$ is the magnetic field, and $R_{0}$ and $R_{s}$
are the ordinary and the extraordinary Hall coefficients,
respectively. By measuring the HE in our films at a temperature
where $R_{s}$ vanishes \cite{rs} we determined $R_{0}$, which
enabled to extract $\mu_{0}R_{s}M_{\bot}$ (where $M_{\perp}$ is
the component of $\mathbf{M}$ which is perpendicular to the film
plane) at all temperatures. This is not sufficient, however, since
we need to determine both components of $\mathbf{M}$. For that we
note that based on symmetry considerations we may assume that if a
field $\mathbf{H}$, that is applied in the $(001)$ plane at an
angle $\phi$ relative to the easy axis, creates a magnetization
$\mathbf{M}$ pointing at an angle $\alpha$ relative to the easy
axis, then applying the same field at an angle $-\phi$ will create
the same magnetization, but at an angle $-\alpha$ \cite{demag}. In
our case the easy axis is at $45^{\circ}$ out of the plane thus
symmetry considerations yield that
$M_{\parallel}(\phi)=M_{\perp}(-\phi)$ where $M_{\parallel}$ and
$M_{\perp}$ are the in-plane and perpendicular components of
$\mathbf{M}$, respectively. Consequently, by measuring the Hall
resistivity ($\rho_{EHE}$) at $\phi$ and $-\phi$ we obtain:
$\rho_{EHE}\left( \phi\right) =\mu_{0}R_{s}M_{\perp}(\phi)$ and
$\rho_{EHE}\left( -\phi\right) =\mu_{0}R_{s}M_{\parallel}(\phi)$,
which allows to determine $\mathbf{M}$ (multiplied by
$\mu_{0}R_{s}$). Figures \ref{mr}a and \ref{mr}b show the change
in the magnitude and direction of $\mathbf{M}$ as a function of
$\phi$ at $T=170$ K determined with that method. As expected,
$\mathbf{M}$ obtains its maximum value at $\phi=0$ and lags behind
$\mathbf{H}$ except for $\mathbf{H}$ along $a$ and $b$. It seems
that this is the first time that EHE which is sensitive only to
the perpendicular component of the magnetization is used for
extracting the full magnetization vector based on symmetry
consideration. This is possible only due to the tilted easy axis.
Where the easy axis is perpendicular or parallel to the film this
scheme is not applicable.

Figures \ref{mr}(c) and \ref{mr}(d) present the MR measured at
$T=170$ K for $H=6$ T and $8$ T in P$_{ab}$ and P$_{c}$. To fit
the MR data we expand MR in $\mathbf{M}$, noting that due to the
MR symmetry under field inversion the lowest order expansion is of
second order. Since in our experiment $\mathbf{M}$ remains in the
$(001)$ plane it is sufficient to use two components of
$\mathbf{M}$. We use the freedom of choosing the principle axes to
take them in the crystallographic directions of $a$ and $b$.
Therefore, the general
expansion of the MR to lowest order is:$\ $%
\begin{equation}
MR=(\rho(H)-\rho(0))/\rho(0)=A(M_{b}^{2}+\beta M_{a}^{2}+\gamma M_{b}%
M_{a})\label{MR expansion}%
\end{equation}
where $M_{b}$ and $M_{a}$ are the components of $\mathbf{M}$ along
the easy axis ($b$) and hard axis ($a$), respectively. The lines
in Figs. \ref{mr}(c) and \ref{mr}(d) are fits of the MR data based
on the measured $\mathbf{M}$ and Eq. \ref{MR expansion}.

We obtained the following fitting parameters at $T=170$ K:
$\beta=1.4\pm0.2$, $\gamma=-0.6\pm0.1$ for the P$_{ab}$ pattern,
and $\beta=1.7\pm0.2$, $\gamma=-0.1\pm0.1$ for the P$_{c}$
pattern. The error limits indicate the evaluated changes in the
fitting parameters in case there is difference between the
instrumental $\phi$ and the actual $\phi$ of up to $2^{\circ}$.

\begin{figure}[ptb]
\includegraphics[scale=0.45, trim=180 240 200 250]{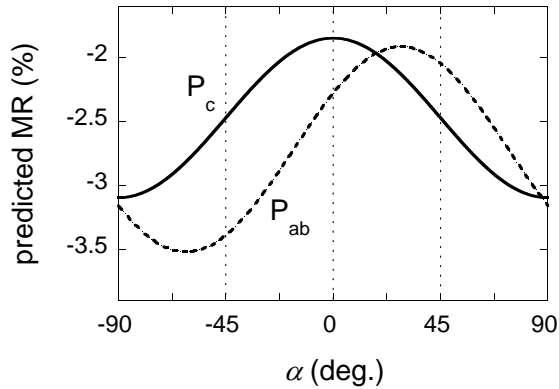}
\caption{The expected MR at $170$ K and $H=8$ T as a function of
the angle $\alpha$ between $\mathbf{M}$ and $b$, assuming $M$ is
constant, for P$_{c}$ (solid curve), and P$_{ab}$ (dashed curve).}
\label{hip}
\end{figure}

The fits of the MR data based on Eq. \ref{MR expansion} allow us
to determine the "clean" AMR effect; namely, how would the
resistivity change if we could rotate $\mathbf{M}$ in the $(001)$
plane without changing its magnitude. Figure \ref{hip} shows the
expected behavior at $T=170$ K for P$_{ab}$ and P$_{c}$ for a
value of magnetization obtained with $H=8$ T along the easy axis.

For P$_{c}$ we note that although $\mathbf{J\perp M}$ there is a
significant AMR ($\beta\neq1$), and that $\gamma=0$ is within our
experimental accuracy, as could be expected from symmetry
considerations. These results exhibit strong dependence of the MR
not only on the angle between $\mathbf{M}$ and $\mathbf{J}$ (which
remains constant) but also on the direction of $\mathbf{M}$
relative to the crystal. For P$_{ab}$ we note that the extremum
values are not obtained for $\mathbf{J\parallel M}$ \ or
$\mathbf{J\perp M}$ but at intermediate angles. In fact, the
extremum\ values for P$_{ab}$ are in between those obtained in
P$_{c}$ (along the $a$ and $b$ axes) and those observed in
polycrystals (parallel and perpendicular to $\mathbf{J}$). This
shows that in our case the AMR related to the orientation of
$\mathbf{M}$ with respect to the lattice is of the same order of
magnitude as the AMR due to the relative orientation of
$\mathbf{M}$ and $\mathbf{J}$. We also note that for
$\alpha=45^{\circ}$, which corresponds to $\mathbf{M}$
perpendicular to the plane the MR is different in P$_{ab}$ and
P$_{c}$ despite the fact that in both cases $\mathbf{J\perp M}$.
This illustrates the dependence of MR on the direction of
$\mathbf{J}$ relative to the crystal.

In conclusion, we have presented AMR investigation of SrRuO$_{3}$
with simultaneous measurements of $\mathbf{M}$ and MR in the same
pattern thus enabling accurate determination of its AMR behavior
despite the change in the magnitude of $\mathbf{M}$ and in its
relative angle with $\mathbf{H}$. The results indicate significant
AMR even in the paramagnetic state, where $\mathbf{M}$ is
relatively small, and large effect of the orientation of
$\mathbf{M}$ and $\mathbf{J}$ relative to the crystal axes.

We acknowledge support by the Israel Science Foundation founded by the Israel
Academy of Sciences and Humanities.


\begin{thebibliography}{9}                                                                                          %
\bibitem {AMR}T. R. McGuire and R. I. Potter, IEEE Trans. Magn.,
\textbf{MAG-11}, 1018 (1975).

\bibitem {easy axis}L. Klein \emph{et al}., J. Phys.: Condens. Matter
\textbf{8}, 10111 (1996).

\bibitem {Marshall}A. F. Marshall \emph{et al}., J. Appl. Phys. \textbf{85},
4131 (1999).

\bibitem {angle}As the
sample is rotated in applied magnetic field, jumps in
magnetoresistance which correspond to magnetization reversal occur
at angles consistent with a uniform direction of the easy axis
throughout the sample.

\bibitem {EHE}J. Smit, Physica \textbf{XXI}, 877 (1955).

\bibitem {rs}L. Klein \emph{et al}., Phys. Rev. B \textbf{61}, 7842 (2000).

\bibitem {demag}Even for a saturated magnetization (which is far
from the case of $170$ K, $6-8$ T), the demagnetizing field is
less than $5\%$ of the fields we apply here.
\end{thebibliography}
\end{document}